\documentclass[aps,prb,twocolumn,superscriptaddress,showpacs]{revtex4-1}
\usepackage{amsmath}
\usepackage[pdftex]{graphicx}
\usepackage[thinspace]{SIunits}

\newcommand{\ud}[1]{{#1^{\dagger}}}

\providecommand \BibitemShut  [1]{\csname bibitem#1\endcsname}%

\begin{document}


\title{A Correlation between the Emission Intensity of Self-Assembled Germanium Islands and the Quality Factor of Silicon Photonic Crystal Nanocavities}

\author{N.~Hauke}\email{hauke@wsi.tum.de}
\author{S.~Lichtmannecker}
\author{T.~Zabel}
\author{F.~P.~Laussy}
\author{A.~Laucht}
\author{M.~Kaniber}\affiliation{Walter Schottky Institut and Physik Department, Technische Universit\"at M\"unchen, Am Coulombwall 4, D-85748 Garching, Germany}
\author{D.~Bougeard}\affiliation{Institut f\"ur Experimentelle und Angewandte Physik, Universit\"at Regensburg, D-93040 Regensburg, Germany}
\author{G.~Abstreiter}\affiliation{Walter Schottky Institut and Physik Department, Technische Universit\"at M\"unchen, Am Coulombwall 4, D-85748 Garching, Germany}\affiliation{Institute for Advanced Study, TU M\"unchen, D-85748 Garching, Germany}
\author{J.~J.~Finley}\affiliation{Walter Schottky Institut and Physik Department, Technische Universit\"at M\"unchen, Am Coulombwall 4, D-85748 Garching, Germany}
\author{Y.~Arakawa}\affiliation{Institute for Advanced Study, TU M\"unchen, D-85748 Garching, Germany}\affiliation{Institute of Industrial Science, Institute for Nano Quantum Information Electronics, The University of Tokyo, 4-6-1 Komaba, Meguro-ku, Tokyo 153-8505, Japan}


\date{\today}

\begin{abstract}
  We present a comparative micro-photoluminescence study of the emission intensity of self-assembled germanium
  islands coupled to the resonator mode of two-dimensional silicon photonic crystal defect nanocavities. The emission intensity is investigated for cavity modes of L3 and Hexapole cavities with different cavity quality factors. For each of these cavities many nominally identical samples are probed to obtain reliable statistics. As the quality factor increases we observe a clear decrease in the average mode emission intensity recorded under comparable optical pumping conditions. This clear experimentally observed trend is compared with simulations based on a dissipative master equation approach that describes a cavity weakly coupled to an ensemble of emitters. We obtain evidence that reabsorption of photons emitted into the cavity mode is responsible for the observed trend. In combination with the observation of cavity linewidth broadening in power dependent measurements, we conclude that free carrier absorption is the limiting effect for the cavity mediated light enhancement under conditions of strong pumping.
\end{abstract}

\pacs{42.60.Da 42.70.Qs 78.55.-m 42.50.Ct}

\maketitle



%


\section{Introduction}

The development of an efficient silicon (Si) based light source is of
great interest in the information technology industry since it would pave the way towards optical interconnects
with CMOS compatibility. If achieved, this major goal would dramatically enhance the signal processing speeds currently achievable in Si-microelectronics. \cite{Jalali07,Jalali06,Soref06,Izhaky06,Kimerling00} Due to
its indirect bandgap light emission in crystalline Si requires the participation of phonons to conserve crystal momentum. Typically, this leads to a very low internal quantum efficiency and Si is, therefore, rarely used as active light emitting material.\cite{Pavesi04} One approach that has been explored to realize an efficient Si based light source is to enhance the material radiative emission efficiency by exploiting cavity quantum electrodynamic effects using photonic crystal (PhC) nanocavities. Enhanced photoluminescence (PL) has been recently reported in crystalline Si PhCs \cite{Iwamoto07,Fujita08,Hauke00} as well as for germanium islands (Ge-islands) embedded in a Si PhC nanocavity.\cite{Li2006,Xia06,Boucaud08,Kurdi08,Xia09} The future development of efficient Si-based light sources and, potentially, even a CMOS compatible laser would revolutionize information technologies. However,
detailed investigations of the nature of the light matter coupling have not been performed, either theoretically or experimentally. Therefore, little is known about the emissive properties of Ge-islands in nanocavities. Furthermore, PhC nanocavities might provide a route to enhance optical activity to a level where optical properties of single Ge-islands can be investigated.

In this article we report on the investigation of the cavity enhanced emission from
self-assembled Ge-islands, which are grown by molecular beam epitaxy (MBE) and are embedded in PhC
nanocavities. We begin by comparing the PL intensities from L3 and Hexapole PhC cavity
modes under comparable conditions of optical pumping as a function of their quality ($Q$)-factors. Averaged over a large number of different and nominally comparable cavities, a very general finding of our work is that  an increase of the average mode PL-intensity  is observed as the mode Q-factor becomes smaller. The average emission intensity saturates for $Q$-factors less than $\approx600$. In order to understand this very clear experimental observation we introduce a cavity quantum electrodynamics (QED) model based on a dissipative master equation for an ensemble of emitters, where each emitter is in a highly excited state to account for the strong optical pumping and weakly coupled to the cavity mode to account for the low oscillator strength. The model exhibits various regimes with strongly different dependencies between mode intensity and $Q$-factor, depending on the spectral emitter ensemble distribution and the emitter-cavity detuning. By comparing this model with the experimentally observed trend we obtain evidence that reabsorption of photons emitted into the cavity mode is responsible for the observed $Q$-factor dependence of the intensity - an effect that ultimately limits the radiative efficiency. By fitting the predictions of our model to our data we extract a photon reabsorption time of $\tau_{\mathrm{abs}}= \unit{2.3\pm0.4}\pico\second$ in very good agreement with the photon reabsorption time $\tau_{\mathrm{abs}}^\star= \unit{1.9\pm0.3}\pico\second$ extracted from power dependent measurements of the cavity mode linewidth. This suggests that free carrier absorption (FCA) takes place under the conditions of strong optical pumping in the cavities. Hence, cavity modes with high $Q$-factors and, thus, long photon lifetimes exhibit a reduced internal radiative efficiency under conditions of strong pumping compared to cavity modes with low $Q$-factors.


\section{Sample structure and experiment}

\begin{figure}
\includegraphics{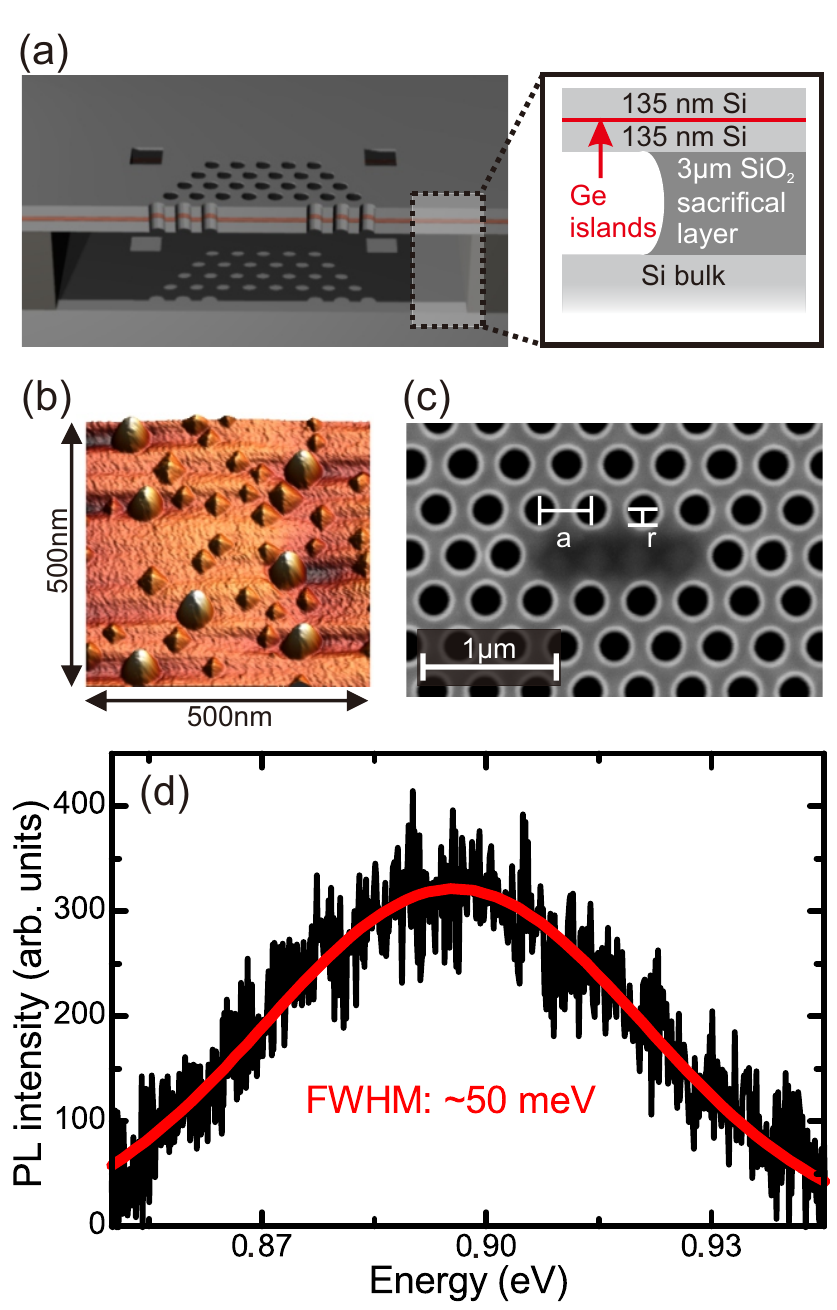}%
\caption{(Color online) (a) Left panel: Schematic cross-sectional representation of the photonic crystal nanocavity structures investigated. Right panel: Layer sequence in the active region.
(b) Atomic force microscope image of the Ge-islands investigated.
(c) SEM image showing a L3 photonic crystal cavity from the top.
(d) $\mu$PL spectrum from Ge-islands in the unpatterned region at $\mathrm{T}=\unit{25}{\kelvin}$. The red line is a Gaussian fit to the data.
 \label{figure1}}
\end{figure}

The sample consists of a 2D PhC that is fabricated into a freely
suspended Si slab, which contains a single layer of self-assembled MBE
grown Ge-islands as illustrated in \mbox{fig. \ref{figure1} (a)}. The
fabrication starts from a silicon-on-insulator (SOI) wafer provided by
Soitec \footnote{\lowercase{w}ww.soitec.com}  with a $\unit{220}{\nano\metre}$ thick crystalline Si layer on top of a $\unit{3}{\micro\metre}$ thick layer of buried SiO$_2$. Before growth, the crystalline silicon layer is thinned to $\unit{50}{\nano\metre}$ using isotropic wet chemical etching with a mixture of 60\% nitric acid (HNO$_3$) and 0.04\% hydrofluoric (HF) acid. After the transfer into the MBE the native oxide is thermally removed by heating to 760$^{\circ}$C for several minutes. The crystal growth is then initialized with a $\unit{85}{\nano\metre}$ Si buffer layer grown at 520-560$^{\circ}$C. Following this, six monolayers (ML) of Ge and a $\unit{135}{\nano\metre}$ thick Si capping layer are deposited at temperatures of 430$^{\circ}$C and 410$^{\circ}$C, respectively. For structural investigations an uncapped layer of islands was grown on the sample surface with the same growth conditions used for the capped island layer. \mbox{Fig. \ref{figure1} (b)} shows an atomic force microscope (AFM) image of the surface islands, revealing a bimodal island distribution with smaller ``pyramids''\cite{Ross1999} and
larger ``domes''\cite{Ross1999}. The emission from the two types of nanostructures is spectrally distinct \cite{Magal2002} and, for the PhC cavities fabricated in this work, the cavity modes are tuned into resonance with the dome emission at $\unit{0.92}{\electronvolt}$. Thus we focus on the interaction between the Ge-dome-islands, which we refer to in the following as ``Ge-islands'', and the PhC nanocavity modes.

After growth, photonic crystal nanostructures were realized using electron beam lithography and subsequent SF$_6$/C$_4$F$_8$ reactive ion etching (RIE) to define hexagonal lattices of air holes with three different periods of $a_1 = \unit{330}{\nano\metre}$, $a_2 = \unit{360}{\nano\metre}$ and $a_3 = \unit{390}{\nano\metre}$. The scanning electron microscope (SEM) image in \mbox{fig. \ref{figure1} (c)} shows a typical PhC with a lattice constant of $a_3 = \unit{390}{\nano\metre}$ containing a L3 PhC nanocavity.\cite{Akahane03} As a final processing step the underlying SiO$_2$ is selectively removed by HF acid to form a freestanding slab membrane.

Optical measurements were performed using a micro-photoluminescence ($\mu$PL) spectroscopy setup. The sample was placed in a liquid helium flow-cryostat for low temperature investigations. To excite the sample we used a continuous-wave (cw) diode-pumped solid-state-laser emitting at $\lambda_{Laser} = \unit{532}{\nano\metre}$, which is focussed by a 100x microscope objective (NA=0.5) to a spot size with a diameter of $\approx \unit{0.8}{\micro\metre}$. The resulting PL signal is collected through the same objective and dispersed by a $\unit{0.32}{\metre}$ imaging monochromator equipped with a 600 lines/mm grating and a liquid nitrogen-cooled \text{InGaAs} linear diode-array.

In \mbox{Fig. \ref{figure1} (d)} we present a typical $\mu$PL spectrum of the Ge-islands emitting in the unpatterned region of the sample. The data, obtained at $\mathrm{T}=\unit{25}{\kelvin}$,  can be fitted well by a Gaussian peak with a full width at half maximum (FWHM) of $\approx\unit{50}{\milli\electronvolt}$ as indicated by the red line.  Due to carrier diffusion our
excitation spot size of $\approx \unit{0.8}{\micro\metre}$ leads to a region with a FWHM of $\approx \unit{1.5}{\micro\metre}$, that generates PL-signal, as obtained by performing $\mu$PL with spatially separated excitation and detection spots. By comparing this finding with
AFM measurements performed on uncapped surface islands, as shown in \mbox{fig. \ref{figure1} (b)}, we estimate that $\approx$ 100 Ge-islands are optically excited by
our laser. Hence, the observed spectrum represents the sum of the individual spectra of $N\approx 100$ single Ge-islands.

\begin{figure}
\includegraphics{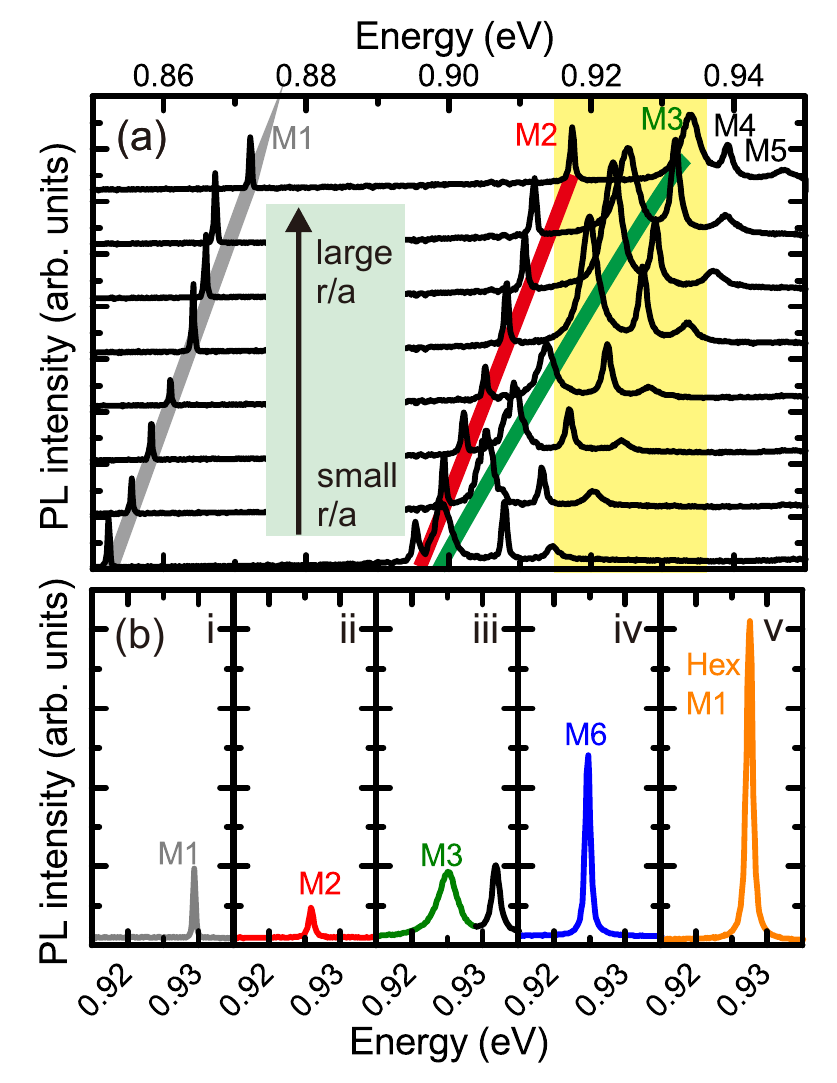}%
\caption{(Color online) (a) $\mu$PL spectra recorded at $\mathrm{T}=\unit{25}{\kelvin}$  from a series of L3 PhC cavities with different air-hole radii, increasing from bottom to top.
(b) Mode emission in the spectral range from $\unit{0.915}{\electronvolt}$ to $\unit{0.935}{\electronvolt}$ [marked yellow in (a)] for different cavity modes: (i) fundamental L3 cavity mode (M1), (ii) first higher energy L3 mode (M2), (iii) second higher energy L3 mode (M3), (iv) fifth higher energy L3 mode (M6) and (v) fundamental dipole mode of a hexapole cavity (Hex M1).
 \label{figure2}}
\end{figure}

In the following, we will compare the emission intensity of cavity modes with various $Q$-factors to study the cavity-emitter coupling. For all experiments reported in this paper the nominal emission intensity was recorded using an optical excitation power density of \mbox{600 kW/cm$^2$} in order to allow a comparative study. Furthermore, it is required that the modes couple spectrally to the same type of islands and, thus, are in a specific spectral window. To do this we tuned the lattice constant of our PhC structures during the fabrication process to coarsly adjust the cavity modes and varied the air hole diameter to fine tune the energy. In \mbox{fig. \ref{figure2} (a)} we plot $\mu$PL spectra recorded at $\mathrm{T}=\unit{25}{\kelvin}$  from a series of L3 PhC nanocavities with a lattice period of $a_2 = \unit{360}{\nano\metre}$ and different air-hole radii, increasing from bottom to top. Six distinct emission lines from the L3 cavity can be observed, five of which are clearly observable in the figure (M1 is the fundamental cavity mode, M2-M5 are the higher energy modes). When increasing the air hole diameter the mode emission shifts systematically to higher energies. For our analysis we will only consider mode emission recorded from samples in the spectral region between $\unit{0.915}{\electronvolt}$ and $\unit{0.935}{\electronvolt}$, highlighted by the yellow shaded region in \mbox{fig. \ref{figure2} (a)}. As can be seen in the figure, we tune the emission of M2 to M5 into this spectral reference region via the fine tuning method alluded to above. For spectral coarse tuning of the mode emission we use the PhCs with lattice constants of  $a_1 = \unit{330}{\nano\metre}$ and $a_3 = \unit{390}{\nano\metre}$ to bring M1 and M6 into the spectral region of interest, respectively.

In order to quantitatively compare the intensity of the PL emission we have to take the mode volume of the nanocavity modes into account. This quantity influences the coupling strength between emitter and cavity and, thus, the cavity mode $\mu$PL intensity. In order to exclude this effect we consider only cavity modes with similar values of the mode volume in our analysis. Therefore, we performed numerical finite difference time domain (FDTD) simulations \footnote{\lowercase{w}ww.rsoftdesign.com} which show that M1, M2, M3 and M6 have comparable mode volumes of $V_{\mathrm{mode}} = 0.65\pm0.09\, (\lambda / n)^3$, whilst M4 and M5 exhibit significantly lower mode volumes of $0.30\, (\lambda / n)^3$ and $0.39\, (\lambda / n)^3$, respectively. Therefore, we excluded M4 and M5 from our analysis and consider only M1, M2, M3 and M6. In addition to the L3 cavities, we also fabricated hexapole cavities \cite{Ryu03} emitting in the spectral reference region and showing a mode volume of $V_{\mathrm{mode}} = 0.63\, (\lambda / n)^3$ for the fundamental dipole mode (M1 hex) of this structure. Hence, we include this mode in the evaluation presented below.
In \mbox{fig. \ref{figure2} (b)} we plot representative mode emission spectra in the spectral region of interest from $\unit{0.915}{\electronvolt}$ to $\unit{0.935}{\electronvolt}$ for a number of different cavity modes: (i) the fundamental L3 cavity mode (M1), (ii) the first higher energy L3 mode (M2), (iii) the second higher energy L3 mode (M3), (iv) the fifth higher energy L3 mode (M6) and (v) the fundamental dipole mode of a hexapole cavity (Hex M1). The PL intensity scale is the same for all graphs.


In addition, we need to take the far field emission profile of the different cavity modes into account, since the far field profile influences the collection efficiency $\eta_{\mathrm{coll}}$ of our optical detection system. As a result, the fraction of light emitted to one hemisphere that is collected by our microscope objective differs for different cavity mode profiles. We obtained the values of $\eta_{\mathrm{coll}}$ for the different cavity modes investigated using FDTD simulations. \mbox{Table \ref{tableModes}} summarizes the simulation results for the cavity modes we include in our analysis. Here, $a$ is the lattice constant used for coarse shifting the mode emission, $V_{\mathrm{mode}}$ the mode volume and $\eta_{\mathrm{coll}}$ the photon collection efficiency.
 \begin{table}
 \caption{Overview of the simulated properties of the PhC cavity modes included in our mode intensity versus $Q$ analysis.  \label{tableModes}}
 \begin{ruledtabular}
 \begin{tabular}{c|c|c|c}
 Mode   & $a$(nm)   & $V_{\mathrm{mode}}(\lambda / n)^3$     & $\eta_{\mathrm{coll}}$  \\ \hline
 M1     & 330       & 0.74                          &  0.197            \\
 M2     & 360       & 0.73                          & 0.108            \\
 M3     & 360       & 0.57                          & 0.437            \\
 M6     & 390       & 0.63                          & 0.468            \\
 M1 hex & 330       & 0.63                          & 0.466            \\

 \end{tabular}
 \end{ruledtabular}
 \end{table}

Finally, we need to exclude measurement errors such as imprecise positioning of the excitation spot on the cavity or varying fabrication quality of different PhCs, all of which would influence the intensity of the PL signal. Hence, we measured 3-7 different photonic crystal nanocavities for each of the mode-types emitting in the spectral window from $\unit{0.915}{\electronvolt}$ to $\unit{0.935}{\electronvolt}$. The mode intensity is extracted by fitting a Lorentzian peak to the spectral profile and averaging the peak areas for modes of the same type. The background stemming from the uncoupled Ge-island emission or from spectrally closely spaced modes (M3 and M4) is subtracted to separate the cavity emission from the background.

\begin{figure}
\includegraphics{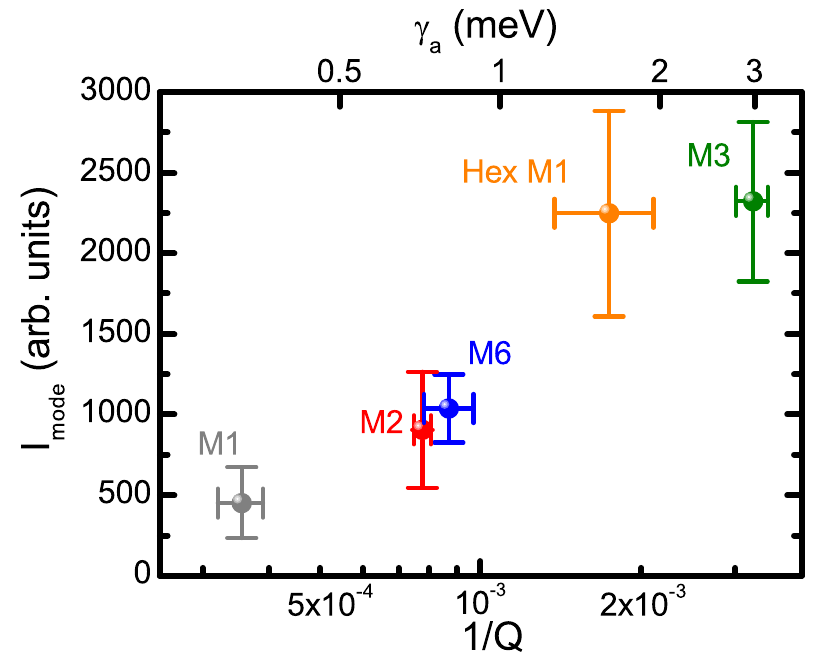}%
\caption{(Color online) Average mode emission intensity as a function of inverse cavity
  $Q$-factor (cavity linewidth $\gamma_a$). Each point is obtained by averaging over several modes
  emitting in the spectral range between $\unit{0.915}{\electronvolt}$ and
  $\unit{0.935}{\electronvolt}$. The error-bars display the
  standard-deviation.
  \label{figure3}}
\end{figure}

In order to obtain the emitted mode intensity we correct the measured intensity
for the mode volume ($V_\mathrm{\mathrm{mode}}$) related change of the number of
islands coupled to the cavity, for the angular collection efficiency
($\eta_{\mathrm{coll}}$) and for the mode degeneracy $D$.\footnote{The dipole mode
  in a hexapole cavity is twofold degenerate.} In
\mbox{fig. \ref{figure3} (a)} we plot the obtained average emitted
mode intensities ($I_{\mathrm{mode}}$) as a function of inverse $Q$ and cavity mode linewidth $\gamma_a$ ($Q=\omega_{cav}/\gamma_a$).
The error bars account for the standard deviation of the intensity and
$Q$-factor distribution, since each point is the average of several
independent measurements performed on a number of cavities. The average emission intensity of cavities with $\gamma_a > \unit{1.5}{\milli\electronvolt}$ ($Q < 600$) are similar. In contrast, for  $\gamma_a < \unit{1.5}{\milli\electronvolt}$ ($Q > 600$), we observe a progressive and systematic decrease in $I_{\mathrm{mode}}$ with decreasing cavity linewidth (increasing $Q$-factors). For the fundamental mode of the L3 cavity (M1), which is the mode with the smallest linewidth, $I_{\mathrm{mode}}$ is strongly reduced to approximately $1/5$ of the intensity of Hex M1 and M3. In order to understand this very clear experimental observation, we introduce in the following section a cavity quantum electrodynamics (QED) model based on a dissipative master equation.


\section{Theory}

\begin{figure}
\includegraphics{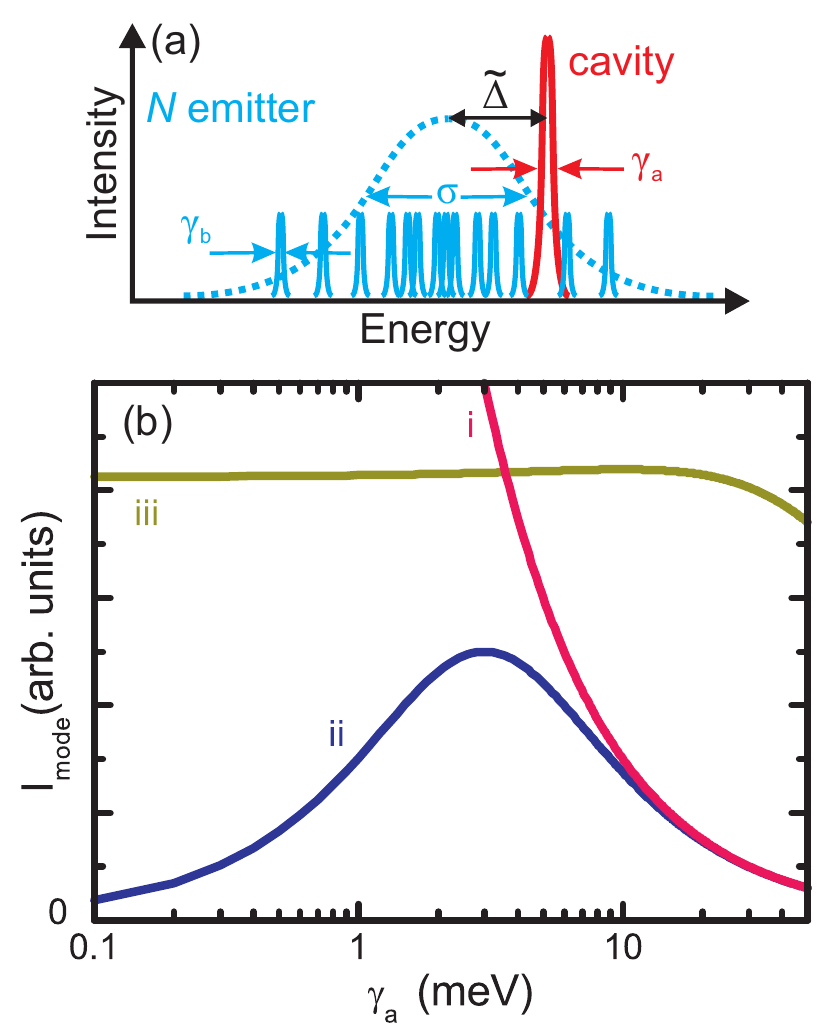}%
\caption{(Color online) (a) Schematic illustration of the parameters used for the theoretical model of $N$ emitters coupled to a cavity: Cavity linewidth $\gamma_a$, emitter linewidth $\gamma_b$, emitter distribution $\sigma$ and effective ensemble cavity detuning $\tilde{\Delta}$.  (b) Theoretical trends in various limiting
  cases of the cavity QED light-matter coupling: (i) Cavity at resonance ($\tilde{\Delta} = 0$) and (ii) at finite detuning ($\tilde{\Delta} = \unit{1.5}{\milli\electronvolt}$) to a spectrally narrow emitter ensemble with a FWHM of $\Gamma_b+\sigma=\unit{0.01}{\milli\electronvolt}$. (iii) Cavity at large detuning ($\tilde{\Delta} = \unit{30}{\milli\electronvolt}$) to a broad emitter ensemble with an effective ensemble linewidth of $\Gamma_b+\sigma=\unit{50}{\milli\electronvolt}$, which corresponds to our system.
  \label{figure4}}
\end{figure}

In this section we develop a cavity-QED model that describes our system of an ensemble of Ge-islands coupled to a cavity mode. Ge-islands are expected to have a very weak transition dipole moment, due to the spatial separation between electrons captured at the Si-Ge interface and the holes localized in the Ge-islands (type II band alignment).\cite{Larsson06} Hence, the coupling parameter $g$ that describes the interaction between emitter and cavity, is expected to be very small. In the following the parameters will be denoted as illustrated schematically in \mbox{fig. \ref{figure4} (a)}, where $N$ is the number of Ge-islands coupled to the cavity mode, $\gamma_b$ is the FWHM of the emission of a single Ge-island, $\sigma$ is the FWHM of the spectral island  ensemble distribution, $\gamma_a$ is the FWHM of the cavity emission and $\tilde{\Delta}$ is the effective spectral detuning between the center of the cavity and the center of the island ensemble. To capture the essential physics that determines the dynamics of such a
system, i.e. $N$ Ge-islands weakly coupled to a cavity mode, we turn
to the simplest possible picture provided by a dissipative master
equation $i\hbar\partial_t\rho=\mathcal{L}\rho$ for the density
matrix~$\rho$ with
$\mathcal{L}\rho=[H,\rho]+\mathcal{L}_\gamma\rho+\mathcal{L}_P\rho$,
where~$H$ is the Hamiltonian that describes the dynamics of the islands-nanocavity system:
\begin{equation}
  \label{eq:FriDec10184015CET2010}
  H=H_\mathrm{free}+\sum_{i=1}^N\hbar g^a_i(\ud{b_i}a+b_i\ud{a})+\sum_{i=1}^N\hbar V_i\ud{b_i}\ud{b_i}b_ib_i\,,
\end{equation}
$H_\mathrm{free}=\hbar\omega_a\ud{a}a+\sum_{i=1}^N\hbar\omega_i\ud{b_i}b_i$
being the free dynamics of the modes and
\begin{equation}
  \label{eq:FriDec10184904CET2010}
  {\mathcal L}_{\gamma_c}\rho=\frac{\gamma_c}{2}(2c\rho\ud{c}-\ud{c}c\rho-\rho\ud{c}c)\,,
\end{equation}
is the Lindblad operator for the cavity modes, where $c=a$, $b_i$ is the decay of the modes. Pumping is included in the same way, as an incoherent source of excitation with Lindblad term
$\mathcal{L}_{P_i}\rho=\frac{P_i}{2}(2\ud{b_i}\rho{b_i}-{b_i}\ud{b_i}\rho-\rho{b_i}\ud{b_i})$. The
nonlinear term $\ud{b_i}\ud{b_i}b_ib_i$ describes phase space filling,
by approximating the real energy level structures of the
island~\cite{Boucaud2001} to be an an equally spaced ladder of
levels~\cite{delvalle08a} with spacing $V_i$ for the i$^{\text{th}}$ dot.  A mean field approximation
$\langle\ud{b_i}b_i\ud{a}b_i\rangle\approx\langle\ud{b_i}b_i\rangle\langle\ud{a}b_i\rangle$
is performed that allows us to truncate the equations of motion
self-consistently.  The main effect of this term is to provide an
effective detuning between the cavity mode and the spectral center of
the emitter ensemble distribution, as well as an
effective broadening of the emitter line. The steady state population
in the cavity
$n_a=\lim_{t\rightarrow\infty}\mathrm{Tr}(\rho(t)\ud{a}a)$ is then
obtained, from which follows the number of photons emitted per unit
time $I_\mathrm{mode}=\gamma_a n_a$. This is the main quantity of
interest in our experiment. A closed form expression can be obtained for a
distribution of islands (a Gaussian distribution is
non-integrable) that shows the combined effect of
detuning~$\Delta=\omega_a-\sum_{i=1}^N\omega_i$ between the cavity
mode and the average position of the islands and their
distribution~$\sigma$:
\begin{equation}
  \label{eq:SunDec12183305CET2010}
 I_\mathrm{mode}=\frac{P_b}{\Gamma_b}\frac{\gamma_a}{(\gamma_a+\Gamma_b+\sigma)}\frac{4g_\mathrm{eff}^2}{\gamma_a}.
\end{equation}
Here, we have retained the leading term only in the coupling strength
$g$, since it is very small in our systems, and have defined the
effective coupling:
\begin{equation}
  \label{eq:SunDec12183545CET2010}
  g_\mathrm{eff}^2=\frac{Ng^2}{1+\left(\frac{\tilde\Delta}{\frac{\gamma_a+\Gamma_b+\sigma}2}\right)^2}\,,
\end{equation}
where~$\tilde\Delta=\sqrt{\Delta^2+[\langle\hbar\sum_i\ud{b_i}b_i
  V_i\rangle/N]^2}$ is the effective detuning that includes the
interactions, and $\Gamma_b$ is the effective broadening of a single emitter that, following Bose statistics, reads
$\gamma_b-P_b$. These two new parameters are however to be considered
the natural and independent ones that describe the system, rather
than the microscopic ones from which they stem. $\Delta$ and
$\gamma_b$ do not play a direct role anymore.

In terms of these effective parameters,
eqn.~(\ref{eq:SunDec12183305CET2010}) is closely related to that of a
single emitter coupled to a cavity (and reduces to it when $N=1$,
$V_i=0$ and $\sigma=0$)~\cite{laussy09a}. The first term in eqn.~(\ref{eq:SunDec12183305CET2010}),
$P_b/\Gamma_b$, is the effective population that builds up among the
islands. The second term, $\gamma_a/(\gamma_a+\Gamma_b+\sigma)$, is the fraction of these excitations which is
available to excite the cavity.  The third term, $4g_\mathrm{eff}^2/\gamma_a$, dominates the dynamics and governs the intensity-linewidth trend.

In fig.~\ref{figure4} (b) we plot eqn.(\ref{eq:SunDec12183305CET2010}) for various limiting
cases of the cavity QED light-matter coupling: (i) Cavity at resonance ($\tilde{\Delta} = 0$) and (ii) at finite detuning ($\tilde{\Delta} = \unit{1.5}{\milli\electronvolt}$) to a spectrally narrow emitter ensemble with an effective ensemble linewidth of $\Gamma_b+\sigma=\unit{0.01}{\milli\electronvolt}$. In curve (iii) we plot eqn.(\ref{eq:SunDec12183305CET2010}) for a cavity at large detuning ($\tilde{\Delta} = \unit{30}{\milli\electronvolt}$) to a broad emitter ensemble distribution with $\Gamma_b+\sigma=\unit{50}{\milli\electronvolt}$. These parameters correspond closely to the expected reality for our system {cf. \mbox{fig. \ref{figure1} (d)}]. For very weak coupling ($g\rightarrow0$), the intensity of the emitted light increases with the quality of the cavity ($\gamma_a\rightarrow0$) if the cavity mode is placed at resonance to a spectrally narrow emitter ensemble ($\tilde{\Delta}=0$). This can clearly be seen by curve (i) in fig.~\ref{figure4} (b). At non-zero detuning with finite spectral
mismatch~$\tilde{\Delta}$ the effective coupling becomes larger with decreasing $Q$ (increasing $\gamma_a$) due to an increasing overlap between the
detuned cavity mode and the collective set of islands, with effective linewidth $\Gamma_b+\sigma$. Hence, making the cavity $Q$ worse gives rise to an increase of the mode emission intensity. For very low $Q$s (large $\gamma_a$) the intensity decreases since the coupling between the emitter ensemble and the cavity gets weak. This is shown representatively by curve (ii) in fig.~\ref{figure4} (b). The trends shown in curves (i) and (ii) can be observed only when the effective ensemble linewidth $\Gamma_b+\sigma$ of the island-ensemble is comparable to the linewidth $\gamma_a$ of the cavity mode. In strong contrast, our system has an ensemble linewidth of $\approx\unit{50}{\milli\electronvolt}$, which is more than an order of magnitude larger than the cavity linewidth. For this case our model predicts that the emitted intensity should exhibit a plateau for the regime of cavity mode linewidths we observe in our experiment ($\unit{0.3}{\milli\electronvolt} < \gamma_a < \unit{3}{\milli\electronvolt}$), quite independently of detuning. This is shown by curve (iii) in fig.~\ref{figure4} (b). Clearly, experimentally we do not observe the theoretically predicted plateau for the high-$Q$ region, but rather a decrease in emission intensity as presented in \mbox{fig. \ref{figure3}}. \footnote{A system in the ``bad emitter regime'' \cite{Pitani2010} would show a decrease of mode intensity with increasing $Q$ since the cavity acts as a spectral filter. However, assuming our system to be in the bad emitter limit leads to unrealistically high values of $g$ (in the order of meV). Hence, we exclude that the intensity decrease we observe is caused by such spectral filter effects.} So far, our cavity-QED model takes only photon emission from the islands into the cavity mode into account, while in the real experiment we have additional effects, such as photon reabsorption, annihilating the photon before it escapes the cavity. With increasing cavity $Q$ the time the photon remains inside the cavity increases and, hence, absorption effects are expected to play an increasingly important role for high $Q$ cavities. This is supported by the fact that M1 is expected to have a high $Q$-factor of $Q_{\mathrm{sim}}\approx 70\,000$ according to our FDTD simulations, but we observe a significantly smaller $Q$-factor in our experiment of $Q_{\mathrm{exp}}\approx 2\,800$.

Due to the spatial separation between electron- and hole-wavefunction and the indirect optical transition in $k$-space we expect the resonant absorption of the Ge-islands to be too weak to explain the $Q$-factor saturation in our experiment. This expectation is supported by the observation that the $Q$-factor of the nanocavity emission is independent of the spectral position relative to the Ge-island ensemble. If reabsorption of photons by Ge-islands would play a dominant role, we would expect to observe an increase of $Q$-factors for cavity modes emitting at the low energy side of the ensemble, since the probability of photon reabsorption decreases there. However, free carrier absorption (FCA) can cause photon reabsorption since we generate a large density of charge carriers in the vicinity of the nanocavity, as we optically excite with a high power density of \mbox{600 kW/cm$^2$}. Such a high optical pumping intensity is required to observe a strong PL signal. We take reabsorption of photons into account as a competing process between the escape of a photon out of a cavity with a photon escape time $\tau_{\mathrm{esc}}$, given by the intrinsic $Q$-factor of the cavity, and the photon reabsorption time $\tau_{\mathrm{abs}}$. The total photon loss time $\tau_{\mathrm{a}}$, given by $1/ \tau_{\mathrm{a}} = 1/ \tau_{\mathrm{esc}} + 1/ \tau_{\mathrm{abs}}$, determines the experimentally observed $Q$-factor and, therefore, the cavity linewidth $\gamma_a$ measured in PL. Hence, the emitted mode intensity $I_\mathrm{mode}^{\mathrm{abs}}$ in the presence of photon reabsorption can be expressed as:
\begin{align}
  \label{eq:absorption}
 I_\mathrm{mode}^{\mathrm{abs}} (\gamma_a)&= I_\mathrm{mode} (\gamma_\mathrm{esc}) \cdot \frac{1/\tau_{\mathrm{esc}}}{1/\tau_{\mathrm{esc}} + 1/\tau_{\mathrm{abs}}} \nonumber \\ &= I_\mathrm{mode} (\gamma_a - \gamma_\mathrm{abs}) \cdot  (1-\frac{\gamma_\mathrm{abs}}{\gamma_a} ) .
\end{align}
$I_\mathrm{mode}^{\mathrm{abs}} (\gamma_a)$ denotes the emitted PL intensity of a cavity with a linewidth of $\gamma_a$ when including absorption, $I_\mathrm{mode} (\gamma_\mathrm{esc})$ denotes the emitted PL intensity of the same cavity without absorption and, hence, a smaller cavity linewidth $\gamma_\mathrm{esc}$ which is solely determined by $1/ \tau_{\mathrm{esc}}$. The last term describes the competing process between the escape of a photon out of the cavity and photon reabsorption, where $\gamma_\mathrm{abs}$ is the cavity linewidth given by the absorption.

\section{Discussion}

In \mbox{fig.~\ref{figure5} (a)} we present the theoretically predicted trend of the mode emission intensity as a function of cavity linewidth $\gamma_a$ for our system in the absence of photon reabsorption [dotted line, cf. fig.~\ref{figure4} (b)].
Furthermore, we plot a fit (full line) of the model to the experimental datapoints (grey, points) from \mbox{fig. \ref{figure3}}, using the same parameters and now including reabsorption of photons as described by eqn. \ref{eq:absorption}. There, we used $\gamma_\mathrm{abs}$ and the normalization of $I_\mathrm{mode}$ as fitting parameters. The theoretically predicted trend follows our experimental data by showing a decrease of emission intensity with decreasing $\gamma_a$ (increasing $Q$-factors). Obviously, we do not expect to observe mode emission for linewidths $\gamma_a$ smaller than $\gamma_\mathrm{abs}$, as highlighted by the red area. From the fit we extract $\gamma_\mathrm{abs} = \unit{0.27\pm0.05}{\milli\electronvolt}$, which corresponds to a photon reabsorption time of $\tau_{\mathrm{abs}}= \unit{2.3\pm0.4}\pico\second$. The presence of FCA is supported by the fact that we observe, with increasing optical pumping power, a broadening of the cavity modes. This is an effect, which El Kurdi \emph{et al.} \cite{Kurdi08} have observed at similar excitation power densities for Si photonic crystal nanocavities with embedded Ge-islands. An example of such pump power dependence investigations for M2 is shown in \mbox{fig.~\ref{figure5} (b)}, where we plot the mode linewidth of M2, $\gamma_a^{\text{M2}}$, as a function of optical excitation power density. In our system, at low optical excitation power densities, the linewidth of M2 stays constant at $\gamma_{\mathrm{esc}}^{\mathrm{M2}} = \unit{0.74\pm0.02}{\milli\electronvolt} $ (marked by the horizontal dotted line), as it is solely determined by the intrinsic photon escape time $\tau_{\mathrm{esc}}$ of the cavity mode. When increasing the excitation power density above \mbox{250 kW/cm$^2$}, we observe a strong increase in linewidth. At \mbox{600 kW/cm$^2$} (marked by the green vertical line), which corresponds to the optical excitation power density we used in our comparative experiment, $\gamma_a^{\text{M2}}$ is increased to $\unit{0.97\pm0.02}{\milli\electronvolt}$. When further increasing the optical excitation power, the absorption coefficient is increased due to heating caused by the excitation laser. This leads to a strong increase of optically generated charge carriers inside the cavity and, as a result, $\gamma_a^{\text{M2}}$ increases strongly.
\begin{figure}
\includegraphics[scale=1.0]{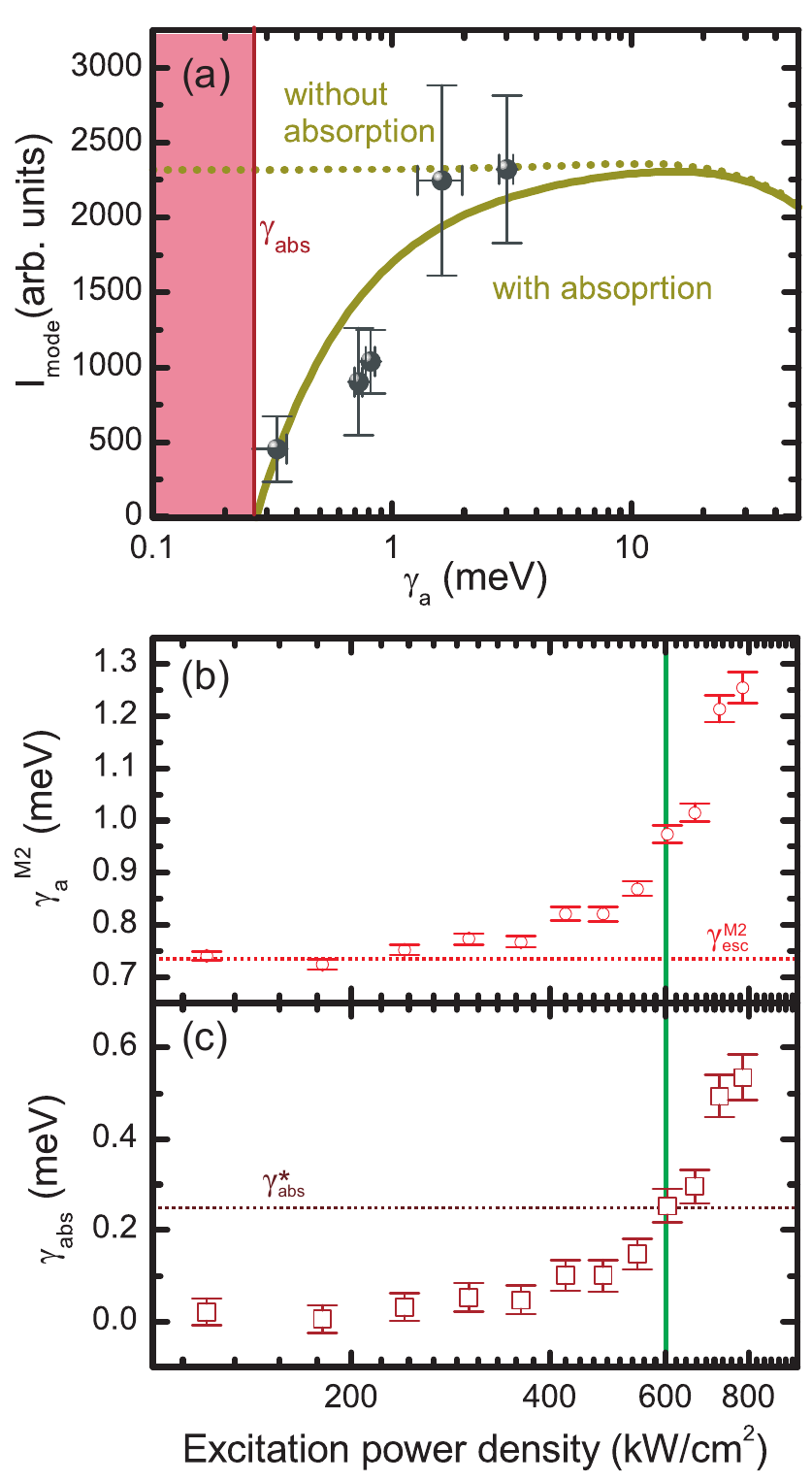}%
\caption{(Color online) (a) Mode emission intensity as a function of cavity linewidth $\gamma_a$: (dotted line) Theoretically predicted trend for our system with an emitter ensemble distribution with  $\Gamma_b+\sigma=\unit{50}{\milli\electronvolt}$ and an ensemble cavity detuning of $\tilde{\Delta} = \unit{30}{\milli\electronvolt}$, in the absence of photon reabsorption and (solid line) including reabsorption of photons with an absorption of $\gamma_\mathrm{abs} = \unit{0.27\pm0.05}{\milli\electronvolt}$ together with the experimental datapoints from \mbox{fig. \ref{figure3}}.
(b) Cavity mode linewidth of M2, $\gamma_a^{\text{M2}}$, as a function of optical excitation power density. The vertical line marks the optical excitation power density we used in our comparative study (\mbox{600 kW/cm$^2$}). The horizontal dotted line marks the intrinsic linewidth $\gamma_{\mathrm{esc}}^{\mathrm{M2}}$ of M2.
(c) $\gamma_\mathrm{abs}$, extracted from $\gamma_a^{\text{M2}}$, as a function of optical excitation power density. The vertical line marks the optical excitation power density we used in our comparative study (\mbox{600 kW/cm$^2$}). The horizontal dotted line marks the extracted absorption linewidth of $\gamma_{\mathrm{abs}}^\star= \unit{0.23\pm0.04}{\milli\electronvolt}$ at this power.
  \label{figure5}}
\end{figure}
Since we can extract the intrinsic linewidth of M2 from the measurements at low power, we can extract the reabsorption $\gamma_\mathrm{abs}$, using $\gamma_\mathrm{abs} = \gamma_a^{\text{M2}} - \gamma_{\mathrm{esc}}^{\mathrm{M2}}$. In \mbox{fig.~\ref{figure5} (c)} we plot $\gamma_\mathrm{abs}$ as a function of optical excitation power density. At $\mbox{600 kW/cm$^2$}$ (marked by the green vertical line), which corresponds to the optical excitation power density we used in our comparative experiment, we obtain $\gamma_{\mathrm{abs}}^\star= \unit{0.23\pm0.04}{\milli\electronvolt}$, corresponding to a photon reabsorption time of $\tau_{\mathrm{abs}}^\star= \unit{1.9\pm0.3}\pico\second$. This is in remarkable agreement with $\tau_{\mathrm{abs}}= \unit{2.3\pm0.4}\pico\second$, as extracted from the fit in $\mbox{fig.~\ref{figure5} (a)}$. Thus, these observations lend strong support to the argument that FCA limits the emission enhancement. A major conclusion of our work is that the overall emitted intensity from optically pumped PhC nanocavities with embedded Ge-islands is mainly limited by FCA, caused by the high charge carrier density from strong optical excitation. Concerning the external quantum efficiency, high-$Q$ cavities are actually performing worse than low-$Q$ cavities due to the increased probability of reabsorption of photons emitted into the cavity.

\section{Conclusion}

We reported a comparative study of PL emission from
PhC nanocavities with embedded Ge-islands by low temperature $\mu$PL
spectroscopy.  First we investigated a number of different L3 and Hexapole PhC cavity modes
with various $Q$-factors. For a valid comparison we considered only
mode emission in a well defined spectral region, corrected for the
mode volume related change of emitters pumping the cavity mode and the
far field radiation pattern. With increasing
$Q$-factors we observed a decrease of the PL-intensity emitted by the cavity mode.  We then introduced a cavity-QED model based on a dissipative master equation
to understand the dynamics of an ensemble of emitters, which are in highly excited states due to strong optical pumping and in very
weak-coupling with the cavity mode. With this model we could identify various regimes of PL intensity versus $Q$-factor trends, depending on the emitter ensemble distribution and the spectral emitter cavity  detuning. By comparing the theoretically predicted trend of our system with the experimental data, we concluded that reabsorption of photons emitted into the cavity limits the emission enhancement via the cavity mode. We extracted a photon reabsorption time of  $\tau_{\mathrm{abs}}= \unit{2.3\pm0.4}\pico\second$ and concluded, by comparing this value to the photon reabsorption time $\tau_{\mathrm{abs}}^\star= \unit{1.9\pm0.3}\pico\second$ extracted from power dependent measurements, that the absorption is caused by free carriers. Therefore, cavity modes with high $Q$-factors and, hence, long photon lifetimes exhibit a reduced radiative quantum efficiency compared to cavity modes with low $Q$-factors.

\section{Acknowledgements}

We would like to thank E. del Valle (University of Southampton, Southampton, United Kingdom) and P. Senellart (Laboratoire de Photonique et Nanostructures, Marcoussis, France)
for fruitful discussions. We acknowledge financial support from the German Excellence Initiative
via the Nanosystems Initiative Munich (NIM), IEF project `SQOD' and the TUM International
Graduate School of Science and Engineering (IGSSE).

\end{document}